\documentclass[twocolumn,prl,amsmath,amssymb,showpacs,superscriptaddress,floatfix]{revtex4-1}
\usepackage{color}
\usepackage{graphicx,epsfig}
\usepackage[hypertex]{hyperref}
\usepackage{srctex}
\usepackage{ulem}
\DeclareMathOperator{\Tr}{Tr}
\DeclareMathOperator{\Real}{Re}\DeclareMathOperator{\Imag}{Im}
\DeclareMathOperator{\Ai}{Ai}\DeclareMathOperator{\Bi}{Bi}
\begin{document}
\author{N. M. Chtchelkatchev}
\affiliation{Institute for High Pressure Physics, Russian Academy of Sciences, Troitsk 142190, Moscow region, Russia}
\affiliation{Department of Theoretical Physics, Moscow Institute of Physics and Technology, 141700 Moscow, Russia}
\author{A. A. Golubov}
\affiliation{Faculty of science and Technology and MESA+ Institute of Nanotechnology, University of Twente, Enschede, The Netherlands}
\author{T. I. Baturina}
\affiliation{Materials Science Division, Argonne National Laboratory, Argonne, Illinois 60439, USA}
\affiliation{A. V. Rzhanov Institute of Semiconductor Physics SB RAS, Novosibirsk, 630090 Russia}
\author{V. M. Vinokur}
\affiliation{Materials Science Division, Argonne National Laboratory, Argonne, Illinois 60439, USA}
\date{\today}

\title{Stimulation of the fluctuation superconductivity by the $\mathcal{PT}$-symmetry}

\begin{abstract}
We discuss fluctuations near the second order phase transition where the free energy has an additional non-Hermitian term.
The spectrum of the fluctuations changes when the odd-parity potential amplitude exceeds the critical value corresponding to
the $\mathcal{PT}$-symmetry breakdown in the topological structure of the Hilbert space of the effective non-Hermitian Hamiltonian.
We calculate the fluctuation contribution to the differential resistance of a superconducting weak link and
find the manifestation of the $\mathcal{PT}$-symmetry breaking in its temperature evolution.
We successfully validate our theory by carrying out measurements of far from equilibrium transport in mesoscale-patterned
superconducting wires.
\end{abstract}

%

\maketitle
An Hermitian character of the Hamiltonian expressed by the condition $H^\dagger=H$ is a cornerstone of quantum mechanics as it ensures that the energies of its stationary states are real. Yet it was discovered not long ago~\cite{Bender1} that the weaker requirement $H^\ddag=H$, where $\ddag$ represents combined parity reflection and time reversal (${\cal PT}$), introduces new classes of \textit{complex Hamiltonians}~\cite{Feshbach} whose spectra are still real and positive~\cite{Bender1,Bender2,Roy,Uwe}. This generalization of Hermiticity opened a new field of research in quantum mechanics and beyond that had been enjoying ever since a rapid growth.

\begin{figure}[t]
\center
  \includegraphics[width=1.0\columnwidth]{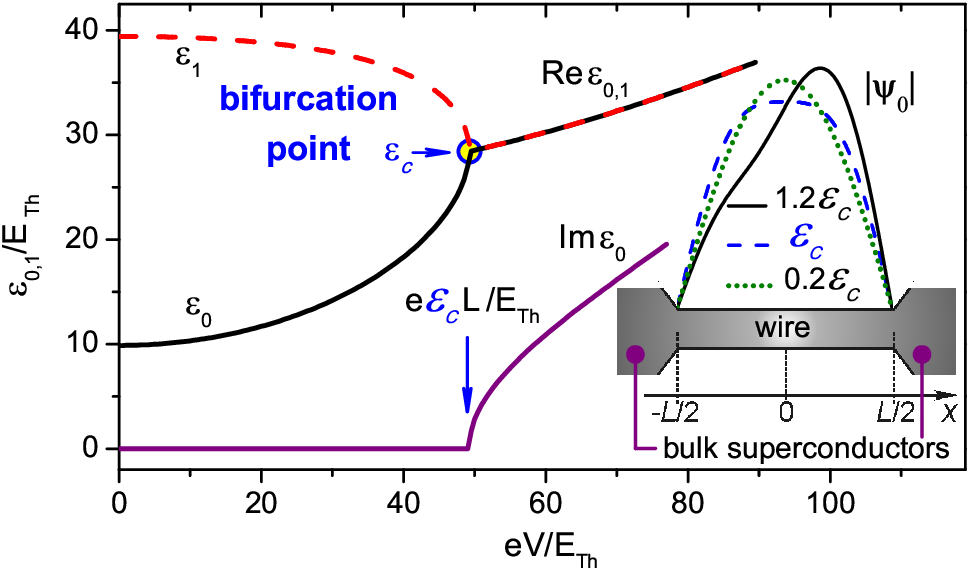}
  \caption{(Color online) Evolution of the two lowest energy levels,
$\varepsilon_{0}$ and $\varepsilon_{1}$ upon applying bias $V={\mathcal E}L$.
At $\mathcal{E}=\mathcal E_c$ the levels merge and form
 the complex conjugate pair at $\mathcal{E}>\mathcal E_c$.
The inset shows the change of the eigenfunction $|\psi_0(x)|$, normalized to unity,
upon variation of $\mathcal E$: $\psi_{0}(x)$ is symmetrical
for $\mathcal{E}\leq\mathcal E_c$ and is asymmetric at $\mathcal{ E}>\mathcal E_c$.
The asymmetry of $\psi_{0}(x)$ is the
signature of the $\mathcal{PT}$-invariance breakdown.}
\label{fig:levels}
\end{figure}

We focus here on the superconducting fluctuations above the superconductor -- normal metal transition in quasi 1D superconducting wire of the finite length $L$ driven far from equilibrium by an electric field $\mathcal E$, see Fig.~\ref{fig:levels}.  We show that either the presence or absence of ${\cal PT}$-symmetry in the Cooperon (fluctuation) propagator, which depends on the magnitude of $\mathcal E$, effects strongly the structure of fluctuations.

 The $\mathcal{PT}$-symmetrical state corresponds to small drive, $|\mathcal E|<\mathcal E_c$, where $\mathcal E_c$ is of the order of the Thouless energy, $E_{\rm Th}=\hbar D/L^2$, the characteristic energy scale of the dirty quasi-one-dimensional conductor, see Fig.~\ref{fig:levels}, where $D$ is the electron diffusion coefficient in the wire and $L$ is its length. This state is nonequilibrium but \textit{stationary} where fluctuating Cooper pairs survive in the presence of the electric field. Breaking the $\mathcal{PT}$-symmetry at $|\mathcal E|=\mathcal E_c$ is the dynamic phase transition from the stationary to the nonstationary dynamic state where the electric field quickly destroys the Cooper-pairs. In this state Cooper pair wave function  qualitatively is represented as a linear superposition of the Ivlev-Kopnin ``kinks'' located at the wire ends~\cite{Kopnin}, having the phases that rotate with the opposite rate. We calculate the fluctuation correction to conductivity and show that for $|\mathcal E|>\mathcal E_c$ this correction is strongly suppressed by an electric field. It implies that $\mathcal{PT}$-symmetry effectively protects Cooper pairs from the detrimental effect of the electric field and stabilizes superconductivity.

%

The dynamics of the superconducting fluctuations is described by the retarded fluctuation propagator $\hat L_R(t,t';x,x')$~\cite{Varlamov_book,chtch}:
\begin{gather}\label{eq:1}
\hat L_R^{-1}=\partial_{\rm t} +{\mathcal H}_{\rm eff}\,, 
\end{gather}
where we use the units $k_{\mathrm B}=e=\hbar=1$. The effective Hamiltonian, $\mathcal H_{\rm eff}[\mathcal E]$,  describes the linearized Ginsburg-Landau (GL) field theory~\cite{GL,Kopninbook}.
In general, $\hat L_R$ can be expanded through the eigen functions
$\psi_n$ and the eigen values $\varepsilon_n$ of ${\mathcal H}_{\rm eff}$:
\begin{gather}\label{eq:L_res}
    \hat L_{R}(\omega;x,x')=\sum_n \frac{\psi_n(x)\psi_n^*(x')}{2i\omega-\varepsilon_n}.
\end{gather}

We show below using the technique developed in Refs.~\cite{Bender1,Uwe,Kopnin}  that $\mathcal{PT}$-symmetry of $\hat L_{R}$ holds at low drives. At large $\mathcal E$ exceeding the certain critical value, $\mathcal E_c$, the $\mathcal{PT}$-invariance breaks down and the two lowest energy states $\varepsilon_0$ and $\varepsilon_1$ of $H_{\rm eff}$ merge. At $\mathcal{E}>\mathcal E_c$ they form the complex conjugate pair (Fig.\ref{fig:levels}). From the general viewpoint of the \textit{catastrophe theory}~\cite{Arnold} bifurcations of $H_{\rm eff}[\mathcal E]$ belong to the so-called fold catastrophe topological class. This class of the bifurcations is (topologically) protected with respect to small local perturbation of ${\mathcal H}_{\rm eff}$
preserving the symmetry of the system. Therefore, in order to establish the existence of the bifurcation and to find its type it would suffice to investigate the
effective Hamiltonian, $\mathcal H_{\rm eff}=-D\nabla_x^2-\tau^{-1}-2i\varphi$. Here $\varphi$ is the potential of the electric field responsible for the nonhermitivity of $\mathcal H_{\rm eff}$.  In application to our problem, neglecting in $\mathcal H_{\rm eff}$ the decay of the
mean-field superconducting order parameter from the reservoirs into the wire does not violate the catastrophe theory classification of bifurcation symmetries.
For the same reason one may choose  the boundary conditions in a form: $\psi(x=\pm L/2\mp0)=\psi(x\to\pm \infty)=0$. Here $\tau$ is the GL-time and $D>0$
is the material constant (e.g., electron diffusion coefficient in the dirty superconductor).
We further discuss the case where $\varphi(x)=\mathcal{E}x$ and $x$ is the coordinate along the wire.

The problem
\begin{gather}\label{eqhpsi}
{\cal{H}}_{\rm eff}\psi=\varepsilon\psi,
\end{gather}
can be solved using the anzatz~\cite{Kopnin}:
\begin{gather}\label{eq:psi}
    \psi(x)=\alpha \Ai(Z)+\beta \Bi(Z),
    \\
    Z(x)=\frac{\varepsilon+2ix\mathcal E}{E_{\mathrm{Th}}}
\left(\frac{E_{\mathrm{Th}}}{2V}\right)^{2/3}\,,
\end{gather}
where $\Ai$ and $\Bi$ are the Airy-functions,  and $\alpha$, $\beta$ are fixed by the boundary conditions.
We absorbed $\tau^{-1}$ into the definition of~$\varepsilon$.
Then the equation determining  the eigenvalues acquires the form:
\begin{gather}\label{eq:eigenvalues}
    F(\varepsilon,\mathcal E)\equiv\Imag\left[\Ai(Z(L/2))\Bi(Z(-L/2))\right]=0.
\end{gather}

The critical field $\mathcal E_{c}$ is the field of emergence of
the first \textit{bifurcation}~\cite{Trenogin,Arnold}
of Eq.\eqref{eq:eigenvalues} corresponding to merging of lowest levels
and is given by the conditions
 \begin{equation}
  F(\varepsilon_c,\mathcal E_c)=0\,,
\,\,\,\,\,\,\,\partial_\varepsilon F(\varepsilon_c,\mathcal E_c)=0\, ,
 \end{equation}
where $\varepsilon_c$ is the value of the energy at the levels merging point.
We find $\mathcal E_c\approx 49.25 E_{\mathrm{Th}}/L$,
where $\varepsilon_c=\varepsilon_0=\varepsilon_1\approx 28.43 E_{\mathrm{Th}}$.
The same conditions give the next bifurcations where higher pairs of levels merge pairwise,
  $\mathcal{E}_c^{(1)}\approx 4 \mathcal{E_{\rm c}}$,
$\mathcal{E}_c^{(2)}\approx 10 \mathcal{E_{\rm c}}$, \textit{etc}...
As we have mentioned above, the bifurcations described here belong to the universality
class of the ``fold catastrophe'' ($A_2$ in ADE classification).
Then $\mathcal E_c$ is the \textit{tipping point} of the catastrophe.

Expanding Eq.\eqref{eq:eigenvalues} near the bifurcation one finds
  $[\frac12(\varepsilon-\varepsilon_c)^2\partial^2_\varepsilon
  +(\mathcal E-\mathcal E_c)\partial_{\mathcal E_c}]F(\varepsilon,\mathcal E)|_{\varepsilon\to\varepsilon_c,\mathcal E\to\mathcal E_c}=0$,
  so,
\begin{gather}\label{eq:bifurcation}
    \varepsilon_{0,1}(\mathcal E)\approx \varepsilon_c\mp E_{\mathrm{Th}}\sqrt{\eta\left[1-\frac{\mathcal E^2}{\mathcal E_c^2}\right]},
    \\\notag
    \eta=\mathcal E_c\frac{\partial_{\mathcal E_c} F(\varepsilon_c, \mathcal E_c)}{E_{\mathrm{Th}}^2\partial^2_\varepsilon F(\varepsilon_c, \mathcal E_c)}\approx
\frac{\pi^2}{\sqrt {2}}\frac{L{\mathcal E_c}}{{E_{\mathrm{Th}}}}.
\end{gather}
The results of the numerical solution of the eigenvalue problem are shown
in Fig.\ref{fig:levels}~\cite{note}. In the limiting case of the semi-infinite wire, $\psi_{0,1}$ for $\mathcal E>\mathcal E_c$ change with coordinates similarly to the solution for the order parameter found in Ref.~\cite{Kopnin}.

Now we proceed with the analysis of the dynamics of the fluctuations in the wire using the following equation:
\begin{gather}\label{eq:LrD}
(\hat L^R)^{-1}\psi=0.
\end{gather}
As long as the field does not exceed the critical value,
$\mathcal{ E}<\mathcal E_c$, the  stationary solution of Eq.\eqref{eq:1}
remains stable and is given by
\begin{gather}\label{eq:2}
\psi(x)\simeq\psi_{0}(x),
\end{gather}
where we have taken $\tau^{-1}=\varepsilon_0$.
This solution is $\mathcal{ PT}$-invariant, i.e.  $|\psi_{0}(x)|=|\psi_{0}(-x)|$,
see Fig.\ref{fig:levels}.
The extremum of $|\psi_{0}(x)|$ is thus located at $x=0$, at the center of the weak link.
The effective field-dependent critical temperature for the superfluid correlations-induced
superfluidity within the weak link is to be found from the relation
$\tau^{-1}=\varepsilon_0$ and is given by
$T_{c}^{\rm (eff)}(\mathcal E)=T_c-\pi\varepsilon_0(\mathcal E)/8$.
The $T_{c}^{\rm (eff)}(\mathcal E)$ dependence become singular near
the critical field  $\mathcal E_c$,
${dT_{c}^{\rm (eff)}}/{d\mathcal E}|_{{\mathcal E}={\mathcal E_c}}=\infty$;
this singularity results in the anomalous behaviour of the nonlinear
fluctuation corrections to the conductivity.
As the field goes above the threshold, $\mathcal E>\mathcal E_c$,
the stationary solution of Eq.\eqref{eq:1} ceases to exist.
The eigenvalues become complex conjugate,  $\Real\varepsilon_0=-\Real\varepsilon_1=\tau^{-1}$
[see inset in Fig.\ref{fig:levels}], and
\begin{gather}\label{eq:23}
    \Imag\varepsilon_0(\mathcal E)=-\Imag\varepsilon_1(\mathcal E)\approx
E_{\mathrm{Th}}\sqrt{\eta\left[\frac{\mathcal E^2}{\mathcal E_c^2}-1\right]}.
\end{gather}
The eigenfunctions at $\mathcal E>\mathcal E_c$ are not $\mathcal{PT}$-invariant any more,
$|\psi_{i}(x)|\neq|\psi_i(-x)|$, $i=1,2$. Thus
\begin{gather}
\psi\sim e^{-i \Imag (\varepsilon_0-\varepsilon_1) t/2}\psi_{0}(x)
+e^{i \Imag (\varepsilon_0-\varepsilon_1) t/2}\psi_{1}(x).
\end{gather}
This implies that the order parameter becomes two-component with the relative phase between the two components rotating with the Josephson frequency $\Imag (\varepsilon_0-\varepsilon_1)$. Since $|\psi_{0}(x)|=|\psi_{1}(-x)|$, the  time averaged order parameter $\langle|\psi(x)|^2\rangle_{\rm time}\sim |\psi_{0}(x)|^2+|\psi_{0}(-x)|^2$
and develops a dip at $x=0$, increasing in amplitude with growing $\mathcal E$. This spot of the relatively suppressed superfluidity finally serves as a heating nucleus in the weak link.

Having calculated the eigenvalues $\varepsilon_n$ and the eigenfunctions, $\psi_n$, $n=0,1,\ldots$, of ${\cal H}_{\rm eff}$, we proceed with the analysis of the superfluidity in the wire under the external drive. We will focus on the effect of superconducting fluctuations on the conductivity of the weak link. The most singular fluctuation contributions to the conductivity come from the Maki-Thompson and Aslamazov-Larkin mechanisms~\cite{Varlamov_book,chtch,varlamov1992}, and the corresponding currents read
\begin{gather}\label{eq:MTAL}
    j^{\scriptscriptstyle{\rm (MT)}}=2DT^2\mathcal E\Tr \{(L^R_{\omega})^\tau (C^R_{\varepsilon})^\tau C^A_{\varepsilon}L^A_{\omega} \}\partial_\epsilon n_F(\tilde\epsilon) ,
    \\\notag
    j^{\scriptscriptstyle{\rm (AL)}}=DT^2\Tr \{(L^R_{\omega})^\tau(C^R_{\varepsilon})^\tau \nabla C^R_{\varepsilon} L^A_{\omega}+h.c.\}\delta n\,,
\end{gather}
where $T$ is the temperature, $\tilde\epsilon=\epsilon+\omega$, $\delta n=n_F(\tilde\epsilon+\varphi)-n_F(\tilde\epsilon-\varphi)$,
$C^{R(A)}_{\varepsilon}=4[D\nabla_{\mathrm x}^2\pm 2 i\varepsilon+\gamma]^{-1}$
is the retarded (advanced) Cooperon propagator, $\gamma$
is the inelastic relaxation rate, and trace `$\Tr$' means the integration
over coordinates, $\varepsilon$ and $\omega$ (the latter two with the weights $1/2\pi$).
Writing Eq.\eqref{eq:MTAL} in terms of $L^{R(A)}$ and $C^{R(A)}$
eigenfunctions and eigenvalues yields the fluctuation correction to the resistance
as (hereafter we restore the physical units and dimensions
of the weak link)
\begin{gather}\label{eq:dsigma}
\delta R\sim -\frac {E_{\rm Th}^2d}{e^2 k_{\mathrm B}TL}\frac1{\sqrt{(\tau^{-1}
-\Real\varepsilon_0(\mathcal E)/\hbar)^2+(\Gamma(\mathcal E)+\gamma)^2}}\,,
\end{gather}
where $d$ is the weak link thickness, $\Gamma=\Imag\varepsilon_0$ and $\tau^{-1}=8(T_{\rm c}-T)/\pi$ while  $T_{\rm c}$ is the critical temperature in the bulk. The resistance displays a pronounced voltage dependence in the range of parameters where either $\hbar/\tau\sim\varepsilon_0(\cal E)$ or ${\cal E}\sim{\cal E}_c$. So, $\delta R(V)$ behaviour can be controlled via changing $\tau$ by cooling or heating the system.

%
\begin{figure}[t]
  \includegraphics[width=1\columnwidth]{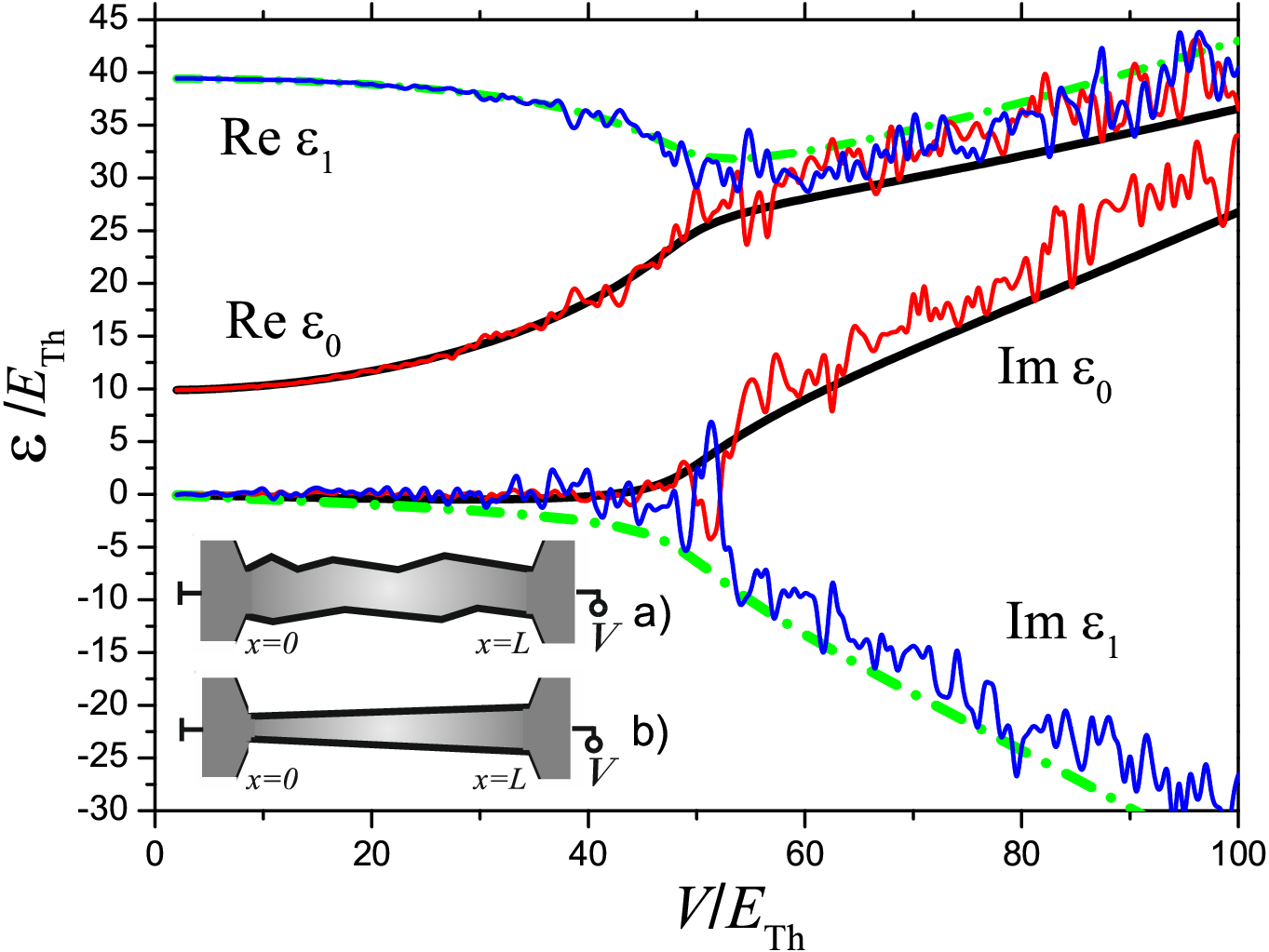}\\
  \caption{Evolution of the two lowest energy levels, $\varepsilon_{0}$ and $\varepsilon_{1}$ upon applying bias $V={\mathcal E}L$ when the wire is not $\mathcal P$-symmetric [so $\mathcal{PT}$-symmetry is also broken] due to the spatial variation of the wire thickness $d$ with $~5\%$ relative amplitude: a) $d$ fluctuates according to the gaussian law, $\varepsilon_0$ is the red curve and $\varepsilon_1$ is the blue one and b) the wire width monotonically changes, $\varepsilon_0$ is the black curve and $\varepsilon_1$ is the green one. In both cases the bifurcation is smoothed and the energies become complex below $\mathcal E_c$. So $\Gamma>0$ and as the result we get the suppression of the fluctuational Cooper pair contributions to the conductivity compared to the case of the $\mathcal{P}$-symmetric wire, see Fig.~\ref{fig:levels}. }
\label{figdisorder}
\end{figure}
%

The regime $|\mathcal E|<\mathcal E_c$ when the system is $\mathcal{PT}$-symmetric favors fluctuational Cooper pairs. When $|\mathcal E|>\mathcal E_c$ the spectrum of fluctuation propagator becomes complex that implies breaking of the Cooper pairs by the electric field with the rate $\tilde \Gamma\neq 0$  that increases with the increase of $\mathcal E$. It follows from Eqs.~\eqref{eq:MTAL} and Eq.~\eqref{eq:dsigma} that then the correction to the resistance from the fluctuating Cooper pairs quickly switches off. The same conclusion follows from the investigation of the superconducting wire where the $\mathcal P$-symmetry is broken due to geometrical imperfection, see Fig.~\ref{figdisorder}.

What we have investigated above was the behavior of the superconducting fluctuations within the framework
of the quadratic Keldysh action describing the fluctuations of the order parameter, see Ref.~\cite{chtch}.
The natural question that arises is whether the revealed bifurcation picture retains in case of large fluctuations
where one has to go beyond the Gaussian approximation.
We expect the affirmative answer
since the predicted instability follows from the symmetry
considerations analogous to those in the general theory of the second phase transitions which, as
one can prove~\cite{chtch}, do not change upon appearance of the higher order terms.
To cast the above reasoning into a mathematical form we note that
on the heuristic level the large fluctuations would result in
modifying~\eqref{eqhpsi} into the nonlinear, but having the same symmetry, Schr\"{o}dinger equation.
The corresponding generalization of~Eq.\eqref{eqhpsi} has the form:
   \begin{equation}\label{Usadel theta}
          \theta ^{^{\prime \prime }}+2i(E+x\mathcal E)\sin \theta =0\,,
    \end{equation}%
The boundary conditions at $x=\pm L/2$ we take in a more general form: $\theta (0)=\theta _{s1}$ and $\theta (L)=\theta _{s2}$, where $\theta_{s1,2}$ are parameterized as follows: $\theta _{s1,2}(E)=\frac{1}{2}(\pi +i\ln \frac{\Delta +E\mp V/2}{\ \Delta -E\pm V/2})$, where $\Delta$ is constant. [For $\mathcal E=0$ Eq.~\eqref{Usadel theta} formally coincides with the Usadel equation \cite{minigap} for the $\theta$ angle parameterizing the quasiclassical retarded Greens function in the superconducting weak link with the order parameter $\Delta$ in the reservoirs.]. Expanding $\sinh\theta$ and identifying $2iE$ with $\epsilon-\tau^{-1}$ and $\theta$ with $\psi$ one recovers~Eq.\eqref{eqhpsi}. We solved Eq.~\eqref{Usadel theta} numerically and found that the first fold-bifurcation appears at $\mathcal E_c\approx 5E_{\mathrm{Th}}/L$ rather than $\approx 49E_{\mathrm{Th}}/L$ found in~Eq.\eqref{eqhpsi}. We thus have demonstrated that even in case of large fluctuations, where the extension beyond the linear approximation is required, the bifurcation of the fluctuation spectrum preserves, while the value of the critical field $\mathcal E_c$ where it occurs may change.

We have focused here on the superconducting wire of relatively small length that generated us the characteristic energy scale  $E_{\rm Th}$. Our solution for the $\mathcal{PT}$-symmetry breaking bifurcation and the fluctuations heavily relied on the discrete nature of $H_{\rm eff}$ spectrum. In the infinite geometry, $L\to\infty$, $E_T\to 0$, and the spectrum of $H_{\rm eff}$ is continuous. Then there is no $\mathcal{PT}$-symmetry breaking bifurcation. Taking the integral in Eq.\eqref{eq:MTAL} over the continues spectrum of $H_{\rm eff}$ we would get fluctuation corrections to the resistance with the form different from Eq.~\eqref{eq:dsigma}. Then the effective pair breaking electric field $\mathcal E_c\sim 1/\xi\tau$~\cite{varlamov1992} as follows from the uncertainty relation between $\tau$ and $\Gamma\sim \xi \mathcal E_c$, where $\xi=\sqrt{D \tau}$.

\begin{figure}[t]
  \includegraphics[width=1\columnwidth]{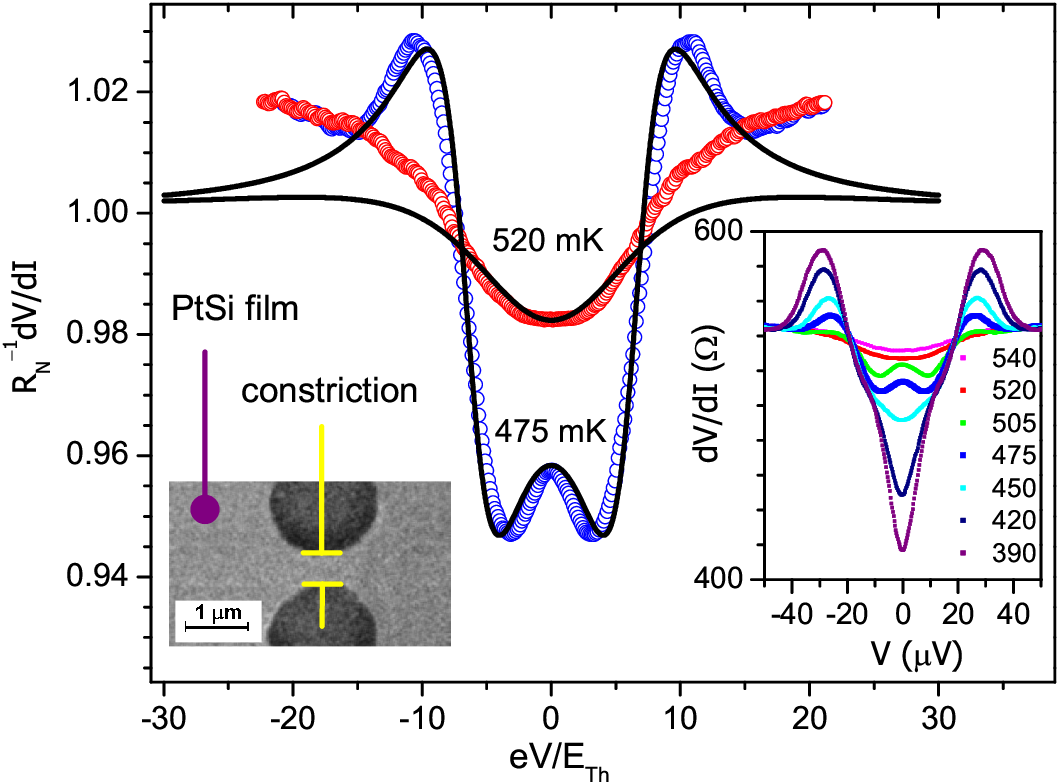}\\
  \caption{The differential resistance of the superconducting weak link. The left inset presents scanning electron microscope image of the experimentally studied system,
the constriction made by the electron beam lithography out of the PtSi film of the thickness $d=6$\,nm and having $T_{\mathrm c}=560$\,mK.
The details of the system preparation are given in~\cite{SNS2}. The right inset displays the full set of the differential resistance vs. $V$ dependences
at different temperatures given in mK in the legend. The central panel shows comparison of two exemplary experimental curves (symbols),
normalized to $R_{\mathrm N}=536\,\Omega$, with the $dV/dI$ calculated from Eqs.\,\eqref{eq:bifurcation} and\,\eqref{eq:dsigma} (solid lines).
We took $L=0.4$\,$\mu$m and $D=6$\,cm$^2/$s~\cite{SNS2}, which
gave $E_{\mathrm{Th}}=2.5\,\mu$eV, and chose $\hbar\Gamma/E_{\mathrm{Th}}=0.176$.
}
\label{fig:delta_sigma}
\end{figure}

In order to test our theory we designed the experiments on mesoscale-patterned ultrathin PtSi wires having small constriction,
as shown in Fig.~\ref{fig:delta_sigma}. The details of the system preparation and parameters of the films are given in \cite{SNS2}
and the Supplementary. The constriction plays the role of a weak link where fluctuation effects are expected to be very strong.
The dimensions and the material characteristics were chosen to create the most favourable conditions for
manifestation of the $\mathcal{PT}$-symmetry breaking effect in the system response to applied voltage bias.
Namely, since the characteristiv energy scale, $E_{\mathrm{Th}}$, is inversely proportional to $L^2$,
the length of the constriction should not be too large in order to diminish the  disguising effect of the thermal broadening.
Another restriction on $L$ is dictated by the condition that the characteristic drive $\mathcal E_c L$
remained less than superconducting gap.
At the same time, in order to suppress Josephson coupling which could prevail over the fluctuation contribution,
one has to take $L \gg \xi_N$,
where $\xi_N=\sqrt{ \hbar D/2\pi k_B T}$ is the decay length for the pair amplitude in diffusive normal conductor.
Taking into account that, according to our calculations, the characteristic energy scale where fluctuations are important is about $10 E_{Th}$,
the above conditions imply that L should not be much larger than 10 $\xi_N$.

Figure 3 shows the  differential resistance, $dV/dI$, of the superconducting weak link as function of the applied voltage bias, $V$.
Upon cooling the system down from the critical temperature, the shape of the measured $dV/dI$-$V$ dependencies
near $V=0$, transforms from the convex one, with the shallow minimum, into the $W$-shaped curve with a peak at $V=0$.
With further decreasing temperature, the central knob inverts, and $ dV/dI(V)$ acquires a pronounced progressively deepening
V-shape developing on top of shallow minimum.
 Importantly, the width of the deep remains equal to that of the maximum (see the curves corresponding to
 $T\leq 450$\,mK in the right inset to Fig.\,3).
 The solid lines in the main panel present the $dV/dI$ vs. $V$ dependences calculated according to
 Eqs.\,\eqref{eq:bifurcation} and\,\eqref{eq:dsigma}, with $\Gamma$ being the only fitting parameter.
 The fit traces perfectly traces the temperature evolution of $dV/dI(V)$, and, most strikingly,
 the $W$-shape at $T=475$\,K in all its details, including maxima in $dV/dI$ at $|eV|\approx 10E_{\mathrm{Th}}$
 and the central knob $5E_{\mathrm{Th}}$ wide.

The similar behaviour of differential resistivity, the evolution from the shallow minimum to maximum and then to the
dip again with the decreasing temperature, was observed in Ref.~\cite{SNS1},
where the quest for the theoretical explanation of this effect was formulated.
Using the parameters given in Ref.~\cite{SNS1} (see Fig. 2 there) we estimate $\xi_N=0.14\,\mu$m at $T=1$\,K,
the bridge length being 2.8\,$\mu$m.
Furthermore, the corresponding $E_{\rm{Th}}\simeq 1.4\mu$V, and one sees that the characteristic voltage of
``saddled'' shaped structure around zero bias
in~\cite{SNS1} is about 40$E_{\rm{Th}}$ in accord with our notion that the $dV/dI$ features develop on the
voltage scale well exceeding Thouless energy.

As a final remark, we stress that the temperature evolution of $dV/dI$ shape results from the confluence of the voltage-dependent fluctuation conductivity, stemming from the Maki-Thompson and Aslamazov-Larkin mechanisms, and the low-voltage quadratic dispersion $[\varepsilon_0(V ) = \varepsilon_0(0)+aV^2$, see Fig.1] of the ground state energy. Importantly, the width of the central knob/peak is $\approx 5E_{Th}$, in a contrast to the more narrow dip in the tunnelling conductivity~\cite{Birge} (the knob in $dV/dI$ corresponds to the groove in $dI/dV$), having the width of $|eV|\approx E_{\mathrm{Th}}$ reflecting the suppression of the electronic density of states by the proximity effect. The observed effect also differs from the zero-bias conductance peak in NS and SNS junctions at low
temperatures~\cite{Klapwijk} originating from the phase-coherent Andreev reflection.

In conclusion, we have demonstrated that the $\mathcal{PT}$-symmetry favors fluctuating Cooper pairs in the superconducting weak link. We have found that the applied electric field exceeding the critical value, $\mathcal E_c$, breaks down the $\mathcal{PT}$-symmetry and destroys the superconducting fluctuations in the weak link and derived the expression for $\mathcal E_c$. Combining effects of superconducting fluctuations and the low-voltage dispersion of the ground state energy of the effective non-Hermitian Hamiltonian of the fluctuating Cooper pairs we have quantitatively described the experimentally observed differential resistance of the weak link in the vicinity of the critical temperature.

We thank N. Kopnin, A. Varlamov, A. Mal'tsev, A. Levchenko, and D. Vodolazov for helpful discussions. The work was funded by the U.S. Department of Energy Office of Science through the contract DE-AC02-06CH11357 and by Russian Foundation for Basic Research (Grants No. 10-02-00700 and 12-02-00152), and the Programs of the Russian Academy of Sciences.

\appendix
\section{Supplementary material}

\subsection{The boundary conditions for GL-equations}
\begin{figure}[hb]
  \includegraphics[width=15mm]{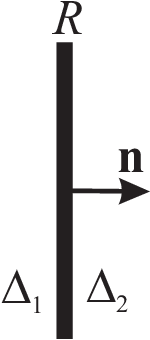}
\caption{The boundary with the normal resistance $R$ between the superconductors with the order parameters $\Delta_1$ and $\Delta_2$.}
 \label{fig:boundary}
\end{figure}
The boundary conditions for GL-equations~\cite{Bcond} on the surface between two superconductors (or a superconductor and a normal metal) can be derived microscopically in the dirty limit using the Usadel equations~\cite{minigap}:
\begin{eqnarray}
\frac{\partial}{\partial x^{\alpha}} \left( {\cal D} \check G
\frac{\partial}{\partial x^{\alpha}} \check G \right) - i[\check H,\check G] =0 ;\\
\label{general}
\check H= \epsilon \hat \sigma_z+ i \hat \sigma_x {\rm Re}(\Delta(x))
+i \hat \sigma_y {\rm Im}(\Delta(x)).
\end{eqnarray}
Here $\Delta$ is superconducting pair potential, ${\cal D}$ stands for diffusivity.
The condition $\check G(\epsilon)$: $\check G^2= \check 1$ holds for the matrix Greens function $\check G$. If a superconductor is divided by an interface,
the boundary conditions for the Greens functions have the form~\cite{Zaitsev}:
\begin{equation}\label{eqs1}
\int dx\, n^\alpha\sigma_1 \check G_1 \frac{\partial}{\partial x^{\alpha}}\check G_1= \sum_n \frac{2 g_0 T_n [\check G_2, \check G_1]_-}
{4+T_n([\check G_2, \check G_1]_+-2)},
\end{equation}
where $\sigma_1$ being specific conductivity in the normal state in the left subspace, $g_0$ is the conductance quantum, $[\ldots,\ldots]_\pm$ is the commutator(anticommutator), $T_n$ are the transmission eigenvalues corresponding to the boundary scattering matrix ($n$ labels the channels) and the indices $1,2$ label the sides of the boundary while $n^\alpha$ is the normal to the boundary.
Derivation of the GL equations from the Usadel equations assumes use of the local equilibrium form of the Greens function. For example, for the retarded components of the Greens function it implies:
\begin{equation}
\hat G_R = \frac{1}{\sqrt{(\epsilon+i0)^2-|\Delta|^2}}\begin{pmatrix}\epsilon & \Delta \cr -\Delta & -\epsilon \cr\end{pmatrix}.
\end{equation}
Taking the retarded component of Eq.~\eqref{eqs1} and expanding it over $\Delta$ to the first order we find the boundary condition for the GL-equations:
\begin{gather}\label{eq:boundary}
\mathbf{N}\cdot\left(\nabla_{\mathbf{r}}-i\frac{2e}{\hbar c} \mathbf{A}\right) \Delta_1 = \frac{\Delta_2-\Delta_1}{\Lambda_1}
\\\notag
\mathbf{N}\cdot\left(\nabla_{\mathbf{r}}-i\frac{2e}{\hbar c} \mathbf{A}\right) \Delta_2 = \frac{\Delta_2-\Delta_1}{\Lambda_2}.
\end{gather}
Here $\Lambda_{1,2} =\sigma_{1,2} S R$, where $R= 1/(g_0 N \langle T\rangle)$ is the interface normal resistance, see Fig.~\ref{fig:boundary}, $S$ is the area of the boundary, $N$ is the number of channels in the boundary, and $\langle T\rangle=\sum_n T_n/N$. Using the estimates: $N\sim (k_F^{(1,2)})^2 S$, $\sigma_{1,2} \sim g_0 (k_F^{(1,2)})^2 l_{1,2}$, we find $\Lambda_{1,2}\sim l_{1,2/}\langle T\rangle$, where $l_{1,2}$ is the mean free path in the left (right) superconductor with the respect to the boundary.

Using this boundary condition in the wire geometry (see Fig.~1 in the paper) we find at $x=\pm L/2$:
\begin{gather}
    \partial_x\Delta(x=L/2-0) = \frac{\Delta_2-\Delta(x=L/2-0)}{\Lambda},
    \\
    \partial_x\Delta(x=-L/2+0) = \frac{\Delta(x=-L/2+0)-\Delta_1}{\Lambda},
\end{gather}
where $\Delta_{1,2}$ denote now the order parameters in the reservoirs.   We can rewrite these boundary conditions as follows:
\begin{gather}
    \partial_x\Delta(x=L/2-0) + \frac{\Delta(x=L/2-0)}{\Lambda}=\frac{\Delta_2}{\Lambda},
    \\
    \partial_x\Delta(x=-L/2+0) -\frac{\Delta(x=-L/2+0)}{\Lambda}=-\frac{\Delta_1}{\Lambda}.
\end{gather}
We can neglect gradients in the boundary condition since $\Lambda$ is of the order of the mean free path $l$ in the wire (the boundaries between the wire and the reservoir are transparent with $\langle T\rangle \approx 1$) that is much smaller than the characteristic scale of the $\Delta(x)$ change with $x$. The connections of the 1D wire and the 3D~(2D)-reservoirs are nonadiabatic. In this case good approximation for the boundary condition is:
\begin{gather}\label{s1}
\Delta(x=\pm L/2\mp0)=\Delta(x\to\pm \infty).
\end{gather}
Approximation of the solution (e.g. a quasiclassical Greens function) at a nonadiabatic connection of the reservoir with the 1D wire by its bulk value comes from the Landauer-Buttiker scattering theory~\cite{Blanter}. This approximation was widely used in superconducting nanostructures, see e.g., Ref.~\cite{Beenakker}, and it was finally refined in the circuit theory~\cite{Nazarov} where the reservoirs were treated as ``nodes''.

\subsection{The boundary conditions for fluctuations}
The relative amplitudes of the superfluid fluctuations and the width of the region near the phase transition from the superfluid to the normal state where fluctuations are relevant are governed by the Ginsburg number $\mathrm{Gi}$~\cite{Varlamov_book}. Typically $\mathrm{Gi}^{\rm(3D)}\ll\mathrm{Gi}^{\rm(2D)}\ll \mathrm{Gi}^{(1D)}$, where $i\mathrm D$, $i=1,2,3$ denotes the effective dimensionality of the sample. So fluctuations in the higher dimensional samples  typically are suppressed compared to fluctuations in the lower dimensional samples. For example, for a dirty superconductor with $k_F l\gtrsim 10$, where $k_F$ is the Fermi wave vector and $l$ is the mean free path, $\mathrm{Gi}^{\rm (3D)}/\mathrm{Gi}^{\rm(1D)}\lesssim 10^{-6}$.  Thus it is reasonable to consider the fluctuations in the bulk of the reservoirs as vanishingly small compared to the fluctuations in the wire. So the eigen functions $\psi$ of $\hat L_{R(A)}$ are localized within the wire.

Material parameters of the wire and the reservoir (including the interaction in the Cooper channel) are identical and the connection between the reservoir and the wire is transparent (no tunnel barrier). On the other hand, the boundary condition for the eigen functions $\psi$ of the fluctuation propagator should be coherent with the boundary conditions for the GL equations. So gradient terms (breaking $\mathcal{PT}$-symmetry) should not not appear in the boundary conditions for $\psi$ at the wire ends $x=\pm L/2$.

The connections of the 1D wire and the 3D~(2D)-reservoirs are nonadiabatic. Then we find boundary condition using the similar arguments we have used above for justification of Eq.~\eqref{s1}:
\begin{gather}
\psi(x=\pm L/2\mp0)=\psi(x\to\pm \infty)=0.
\end{gather}

The question that can be asked at this point is what will change in case of the metallic reservoirs.
The answer is simple if the reservoirs are superconductors but in the normal state and their critical temperatures are close to the critical temperature of the wire material. Then GL theory is valid in the wire and in the reservoirs at the same time. The boundary conditions~\eqref{s1} for the mean-field order parameter $\Delta$ are valid in this case. The gradient terms in the boundary conditions are are again irrelevant. The normal state of the reservoirs does not change the symmetry of the problem. So all our results regarding the bifurcation remain valid in this case.

Situation becomes more complicated when critical temperatures of the reservoirs are much smaller than critical temperature of the wire material. In this case GL-expansion can be used in the wire only. The are no fluctuations in the reservoirs. However, the derivation of the boundary conditions given above is not applicable in this case. Investigation of the role of gradients in the boundary conditions for $\psi$ on $\mathcal{PT}$-symmetry breaking in the wire will be the subject of forthcoming publication.

Another question is related to the influence of the magnetic field on our result. It will enter in the standard way in the gradient terms of $C^{R(A)}$ and the fluctuation propagator. The magnetic field changes the orbital movement of the fluctuating Cooper pairs that is quite difficult in the narrow wire. Therefore, we would expect the orbital effects are only important if the magnetic flux through the bridge crossection exceeds the flux quantum (fields more than 1T). So our results are quite insensitive to the magnetic field.

We have shown in this paper that $\psi_n$ are the building blocks for the fluctuations. Important and difficult question is whether $\psi_n$ have something to do with superconducting (superfluid) phase on the mean field level. The mean-field fully nonequilibrium investigation of superconductivity in the short superconducting wire under external drive have been made in Refs.~\cite{Vodolazov,Pekola}. Based on the results of Refs.~\cite{Vodolazov,Pekola} we can conclude that there exists some characteristic electric field $\tilde {\mathcal E_c}$ where the topological structure of the coordinate dependence of the order parameter strongly changes. We could identify this field with $\mathcal{PT}$-symmetry breaking field $\mathcal E_c$ since the coordinate dependence of the order parameter in Refs.~\cite{Vodolazov,Pekola} qualitatively follows our $\psi_0$, $\psi_1$ below and above $\tilde{\mathcal E_c}$. We will investigate this issue in the forthcoming paper.

\subsection{Sample parameters}

The original polycrystalline superconducting PtSi film with thickness of 6 nm was formed on
the Si substrate. The film had the critical temperature $T_c$=0.56 K. The resistance per square was 104 $\Omega$.
The carrier density obtained from Hall measurements was $7 \times 10^{22}$ cm$^{-3}$, corresponding to the mean-free path
$l$=1.2 nm and the diffusion constant $D$=6 cm$^2$/s estimated using the simple free-electron model ~\cite{Baturina}.
A square lattice of holes was patterned by means of the electron lithography and the subsequent plasma etching,
in order to obtain the structure consisting of islands of the film, with the characteristic dimension
1.3 $\mu$m, connected by narrow necks having the width of 0.4 $\mu$m,
where superconductivity is suppressed by the applied current.
As a result, the investigated sample is an array of the SNS junctions
having superconducting and normal regions made from the same material and therefore no extra resistance originating from the SN interfaces.


\begin{thebibliography}{99}


\bibitem{Bender1}
C.\,M.~Bender and S.~Boettcher, Phys. Rev. Lett. \textbf{80}, 5243 (1998).

\bibitem{Feshbach} H. Feshbach, Ann. Rev. of Nucl. Science \textbf{8}, 49 (1958); E. A. Solov'ev, Sov. Phys. Usp. \textbf{32}, 228 (1989); H. Nakamura in \textit{Nonadiabatic Transition: Concepts, Basic Theories, and Applications} (World Scientific, Singapore, 2002); O. I. Tolstikhin, \textit{et al}., Phys. Rev. A \textbf{70}, 062721 (2004).

\bibitem{Bender2}
C.\,M.~Bender, D.\,C.~Brody, and H.\,F.~Jones, Phys. Rev. Lett. \textbf{89}, 270401 (2002); \textit{ibid} \textbf{92}, 119902 (2004).

\bibitem{Roy} B. Roy and P. Roy, Phys. Lett. A \textbf{359}, 110 (2006).

\bibitem{Uwe} U. G\"{u}nther, \textit{et al}., J. Math. Phys. \textbf{46}, 063504 (2005); J.~Rubinstein, \textit{et al}., Phys. Rev. Lett. \textbf{99}, 167003 (2007).

\bibitem{Kopnin} B. I. Ivlev, N. B. Kopnin, and L. A. Maslova, Sov. Phys. JETP \textbf{56}, 884 (1982);  B. I. Ivlev, N. B. Kopnin, Sov. Phys. Usp. \textbf{27}, 206 (1984).

\bibitem{GL} L. D. Landau and I. M. Khalatnikov, Dokladii Academii Nauk
CCCP 96, 469 (1954); V.L. Ginzburg, and L. D. Landau, Zh. Eksp. Teor. Fiz. \textbf{20}, 1064 (1950).

\bibitem{Kopninbook} N. B. Kopnin, \textit{Theory of Nonequilibrium Superconductivity} (Clarendon Press, Oxford, 2001).

\bibitem{GP} F. Dalfovo, S. Giorgini, L.P. Pitaevskii, \textit{et al}., Rev. Mod. Phys. \textbf{71}, 463 (1999).

\bibitem{Varlamov_book} A.\,I.~Larkin and A.\,A.~Varlamov, \textit{Theory Of Fluctuations In Superconductors}, (Clarendon Press, Oxford, 2005).

\bibitem{chtch} A. Petkovic, N.M. Chtchelkatchev, \textit{et al}., Phys. Rev. Lett. \textbf{105}, 187003 (2010); A. Petkovic, \textit{et al}.,
Phys. Rev. B \textbf{84}, 064510 (2011); N.M. Chtchelkatchev, \textit{et al}., Euro Phys. Lett. \textbf{88}, 47001 (2009).

\bibitem{Arnold} V.\,I. Arnold, \textit{Catastrophe Theory}, Springer, Berlin, 1992; T. Poston and I. Stewart, \textit{Catastrophe: Theory and Its Applications} (Dover, New York, 1998).







\bibitem{Trenogin}
M.\,M. Vainberg and V.\,A. Trenogin,
\textit{Theory of branching of solutions of non-linear equations}
(Noordhoff International, Leyden, 1974).

\bibitem{note} Replacing $\varepsilon_c$ in Eq.\,\eqref{eq:bifurcation} by $\tilde\varepsilon_c(\mathcal E)=\bar\varepsilon_{01}+(\varepsilon_c-\bar\varepsilon_{01})(\mathcal{E}/\mathcal{E} _c)$, where $\bar\varepsilon_{01}=[\varepsilon_0(0)+\varepsilon_1(0)]/2$ provides very accurate approximation of $\varepsilon_{0,1}(\mathcal E)$ in the whole range of $\mathcal E$.

\bibitem{SNS2}
Z.\,D.~Kvon \textit{et al.}, Phys. Rev. B \textbf{61}, 11340 (2000);
T.\,I.~Baturina \textit{et al.}, Phys. Rev. B \textbf{63}, 180503(R) (2001); T.\,I.~Baturina \textit{et al.}, JETP Lett. \textbf{75}, 326 (2002);
T.\,I.~Baturina \textit{et al.}, JETP Lett. \textbf{81}, 10 (2005).

\bibitem{Birge}
S. Gu\'eron \textit{et al.},
Phys. Rev. Lett. \textbf{77}, 3025 (1996).

\bibitem{SNS1}
J. Kutchinsky \textit{et al.},
Phys. Rev. Lett. \textbf{78}, 931 (1997).

\bibitem{Klapwijk}
T.M. Klapwijk,
Physica B \textbf{197}, 481 (1994).

\bibitem{minigap} K.\,D.\,Usadel, Phys. Rev. Lett. \textbf{25}25, 507 (1970).

\bibitem{varlamov1992}
A. A. Varlamov \textit{et al}., Phys. Rev. B \textbf{45}, 1060 (1992);
I. Puica, and W. Lang, Phys. Rev. B \textbf{68}, 054517 (2003).

\bibitem{Bcond} E.~A. Andryushin, V.~L. Ginzburg and A.~P. Silin, Phys.-Usp. \textbf{36}, 854 (1993).

\bibitem{Zaitsev} A. V. Zaitsev, Sov. Phys. JETP 59, 1163 (1984); M. Yu. Kupriyanov and V. F. Lukichev, Zh. Eksp. Teor.
Fiz. \textbf{94}, 139 (1988) [Sov. Phys. JETP \textbf{67}, 1163 (1988)]; Yuli V. Nazarov, Superlattices and Microstructures
\textbf{25}, 1221 (1999).

\bibitem{Blanter} Ya. M. Blanter, and  M. Buttiker, Phys. Rep. \textbf{336}, 1 (2000).

\bibitem{Beenakker} C. W. J. Beenakker and H. van Houten, Phys. Rev. Lett. \textbf{66}, 3056 (1991); C. W. J. Beenakker, Rev. Mod. Phys. \textbf{69}, 731 (1997).

\bibitem{Nazarov} Yuli V. Nazarov, Superlattices and Microstructures \textbf{25}, 1221 (1999).

\bibitem{Vodolazov}
A.\,K.~Elmurodov \textit{et al.},
Phys. Rev. B \textbf{78}, 214519 (2008).

\bibitem{Pekola} N. Vercruyssen, T. G. A. Verhagen, M.G. Flokstra, J. P. Pekola, and T. M. Klapwijk, Phys. Rev. B \textbf{85}, 224503 (2012).

\end{thebibliography}
\end{document}